# Physical Characterization of ~2-meter Diameter Near-Earth Asteroid 2015 TC25: A possible boulder from E-type Asteroid (44) Nysa

Authors: Vishnu Reddy, Juan A. Sanchez, William F. Bottke, Audrey Thirouin, Edgard G. Rivera-Valentin, Michael S. Kelley, William Ryan, Edward A. Cloutis, Stephen C. Tegler, Eileen V. Ryan, Patrick A. Taylor, James E. Richardson, Nicholas Moskovitz, Lucille Le Corre

**Abstract**

Small near-Earth asteroids (>20 meters) are interesting because they are progenitors for meteorites in our terrestrial collection. Crucial to our understanding of the effectiveness of our atmosphere in filtering low-strength impactors is the physical characteristics of these small near-Earth asteroids (NEAs). In the past, characterization of small NEAs has been a challenge because of the difficulty in detecting them prior to close Earth flyby. In this study we physically characterized the 2-meter diameter near-Earth asteroid 2015 TC25 using ground-based optical, near-infrared and radar assets during a close flyby of the Earth (distance 69,000 miles) in Oct. 2015. Our observations suggest that its surface composition is similar to aubrites, a rare class of high albedo differentiated meteorites. Aubrites make up only 0.14 % of all know meteorites in our terrestrial meteorite collection. 2015 TC25 is also a very fast rotator with a rotation period of 133 seconds. We compared spectral and dynamical properties of 2015 TC25 and found the best candidate source body in the inner main belt to be the 70-km diameter E-type asteroid (44) Nysa. We attribute difference in spectral slope between the two objects to the lack of regolith on the surface of 2015 TC25. Using the albedo of E-type asteroids (50-60%) we refine the diameter of 2015 TC25 to 2-meters making it one of the smallest NEA ever to be characterized.

## 1. INTRODUCTION

Small bodies in the solar system are time capsules that have recorded the conditions during planet formation. Studying these objects will not only help us better understand how our planet formed but also how large impacts help shape the course of life on Earth. The recent Chelyabinsk bolide event (Boslough 2013) has reminded us of the threats posed by small NEAs (<20 meters in diameter) to population on Earth. With an estimated population size of 8-9 million (Harris and D'Abramo, 2015), objects the size of the Chelyabinsk impactor are routinely discovered by surveys prior to their flybys through Cislunar space each lunation. However, due to their small size, optimal geometry and observing windows for ground-based characterization are rare. Ground-based telescopes have characterized only a handful of NEAs smaller than 20 meters (mostly photometry and visible wavelength spectroscopy). Here we present results from our characterization of a 2-meter diameter NEA using optical, infrared and radar telescopes to constrain its rotation period, composition and source body in the main asteroid belt.

## 2. PHOTOMETRIC STUDY

*2.1 Observations and Data Reduction*

We present data obtained at the 4.3-m Lowell Discovery Channel Telescope (DCT), the 3-m NASA Infrared Telescope Facility (IRTF), and the 2.4m Magdalena Ridge Observatory (MRO) telescope. All data were taken within one day of discovery on 12 October 2015. The DCT images were obtained using the Large Monolithic Imager (LMI), which is a 6144×6160 pixel CCD. The total field of view is 12.5'×12.5' with a pixel scale of 0.12"/pixel (unbinned). Images were obtained using the 2×2 binning mode. Broad-band VR-filter has been used to maximize the signal-to-noise of the object. Exposure time was 3s. The IRTF data were obtained using MIT Optical Rapid Imaging System (MORIS), a 512x512 pixel Andor CCD co-mounted with SpeX was used. MORIS has a pixel scale of 0.11"/pixel and a field of view of 1' square. The MRO images were obtained using a back-illuminated CCD with a pixel scale of 0.53"/pixel and a field of view of 4.5'×4.5'. VR filter and an exposure time of 10s have been used to maximize the signal-to-noise of the object. Observational circumstances are provided in Table 1.

DCT and IRTF datasets were calibrated and reduced using the techniques described in Thirouin et al. (2014). The IRAF ccdproc and apphot routines were used to reduce the MRO data and extract instrumental photometry. Times-series photometry was inspected for periodicities by means of the Lomb technique (Lomb 1976) as implemented in Press et al. (1992).

*2.2 Lightcurve Results*

2015 TC25 is one of the smallest NEOs observed for lightcurve with an absolute magnitude (H) of 29.5 (corresponding to a diameter of 2-m for a E-type, see Section 3.2). Only 4 more objects with H>29 have a lightcurve: 2006 RH120 with a period of 2.75 min (Kwiatkowski et al. 2008), 2008 JL24 rotates in 3.23 min (Warner et al. 2009a), 2010 WA in 0.51 min and 2012 XB112 with a rotation of ~2.6 min[1]. From its lightcurve we determined that 2015 TC25 is also a fast rotator with a rotational period of about 2.23 min (133.8 seconds) (Figure 1). The composite lightcurve displays an amplitude of 0.40 mag with the peaks asymmetric by approximately 0.08 mag, suggesting that the object has an irregular shape.

Lightcurve amplitude can be corrected from phase angle effect (Equation in Section 5 in Zappala et al. 1990), i.e we can estimate the lightcurve amplitude at a phase angle of 0°. Phase angle ranged from 25° to 32° during our observing (average of ~29°). The lightcurve amplitude of 2015 TC25 corrected from phase angle is 0.24 mag using a slope of 0.023 deg$^{-1}$ (Zappala et al. 1990).

---

[1] http://infohost.nmt.edu/~bryan/research/work/neo/lightcurves/

Assuming 2015 TC25 as triaxial ellipsoid with axes a>b>c and rotating along the c-axis, the lightcurve amplitude varies as a function of the viewing angle, and we can estimate the object elongation (a/b) (Binzel et al. 1989). Using Equation 4 of Kwiatkowski et al. (2010), we estimated an a/b ratio ~ 1.22 which suggests a moderate deformation of the object.

## 3. CONSTRAINING COMPOSITION

*3.1 Radar Observations*

Asteroid 2015 TC25, with absolute magnitude $H$=29.5, is one of only three asteroids with $H$>29 ever detected at Arecibo Observatory. Arecibo S-band (2380 MHz, 12.6 cm) radar observations of 2015 TC25 consisted of transmitting a circularly polarized, monochromatic tone for approximately the time taken for light to reach the asteroid, reflect, and return to the 305-m antenna followed by reception of the echo in both the same-circular (SC) and opposite-circular (OC) polarizations for a similar amount of time. Reflection from a plane mirror would result in a purely OC echo, while realistic surfaces introduce a degree of depolarization. Hence, the polarization ratio (SC/OC) depends upon the scattering properties of the surface and near-surface material and in some cases can differentiate between taxonomic classes, *e.g.*, between the S complex and E-type asteroids (Benner et al. 2008).

Radar observations were conducted on 2015 October 17 and 18 when 2015 TC25 was an average of four times the separation between the Earth and Moon. Observational circumstances for radar data are provided in Table 1. A total of 72 transmit-receive cycles were completed on the first day while 109 were completed on the second. An update to the observing ephemeris was implemented between the tracks, adjusting the Doppler frequency by 0.20 Hz compared to the pre-experiment three-sigma uncertainty of 26 Hz, a minor 0.02-sigma correction. The polarization ratio SC/OC, determined as the quotient of the integrated SC and OC echo powers in Figure 2, was 1.10 ± 0.23 and 0.77 ± 0.15 (one-sigma) on October 17 and 18, respectively. Combining the observations suggests a polarization ratio of ~ 0.9 with a possible range of 0.62 – 1.33. These values are significantly elevated compared to S- and C-complex asteroids (Benner et al. 2008) with mean polarization ratios of 0.27 and 0.29, respectively, but are consistent with the range of polarization ratios measured for known E-type asteroids of 0.74 – 0.97.

Also note in Figure 2 that upon reflection the monochromatic transmitted tone is broadened by the asteroid's rotation according to

$B = (4\pi D/\lambda P) \cos \delta$,

where $B$ is the measured echo bandwidth, $D$ is the diameter of the asteroid, $\lambda$ is the transmitted wavelength, $P$ is the rotation period of the asteroid, and $\delta$ is the sub-radar latitude. Considering the measured period from lightcurves of 133.8 s and an estimated diameter of 2 m, producing the observed bandwidth *B* of 0.8 Hz implies the sub-radar latitude during the observations was ~60°. Larger diameters would require a nearly pole-on geometry, while an equatorial view allows for a diameter of 1 m.

*3.2 Spectroscopic Observations*

Near-infrared spectroscopic observations of 2015 TC25 were obtained remotely using the SpeX instrument in prism mode on NASA IRTF (Rayner et al. 2003) on 12 October 2015 between 8:22-9:35 UTC when the asteroid was V. mag 17.5 and a phase angle of 32° (Table 1). Apart from the asteroid, local G-type star SAO167240 and solar analog star SAO31899 were also observed for telluric and solar slope corrections. The prism data were processed using the IDL-based Spextool provided by the NASA IRTF (Cushing et al., 2004). Figure 3 shows the spectrum of 2015 TC25 with a weak absorption band at 0.9 µm and a negative spectral slope (decreasing reflectance with increasing wavelength) beyond 1.1 µm. Band I center and Band I depth were measured following the protocols described in Cloutis et al. (1986). We found that 2015 TC25 has a Band I center of 0.905±0.003 µm, and a Band I depth of 5.4±0.2%. Due to their relatively small heliocentric distance, low albedo NEAs show a distinct upturn in reflectance (shorter wavelength end of the Planck curve) in near-IR wavelengths beyond 2.0-µm (Rivkin et al. 2005; Reddy et al. 2012). 2015 TC25 does not show any evidence of such an upturn suggesting a moderate-high surface albedo consistent with E-type asteroids.

*3.2.1 Relationship with E-type asteroids*

The moderate-high albedo and the radar SC/OC ratio of 2015 TC25 suggest that this object belongs to the E-type asteroids, which are characterized for having the highest albedos (~0.4-0.6) among all known asteroid classes (Zellner and Tholen 1985; Tholen, 1989). E-type asteroids in the main belt are mainly concentrated in the Hungaria region at ~1.9 AU, although some of them have been also found dispersed throughout the inner part of the main belt from ~2.1 to 2.7 AU. These objects have been traditionally linked to the enstatite achondrite meteorites (aubrites) based on their high albedos and spectral shape (e.g., Zellner et al. 1977; Cloutis et al. 1990; Gaffey et al. 1992; Burbine et al. 2002). Examples of aubrite spectra are shown in Figure 4.

Gaffey and Kelley (2004) determined that there are at least three spectrally distinct subtypes (designated as E(I), E(II) and E(III)) among the E-type asteroids. The E(I) subtypes lack any discrete diagnostic absorption feature in the VIS-NIR wavelength range. The E(II) exhibit a relatively strong feature centered near 0.5 µm and some times a weaker feature near 0.96 µm. These features are attributed to the presence of oldhamite (CaS). The E(III) subtypes exhibit a well-defined ~0.9 µm absorption feature due to enstatite pyroxene containing traces of $Fe^{2+}$. The feature at 0.5 µm seen in the E(II) subtypes is not present in the spectra of E (III).

Clark et al. (2004), on the other hand, divided the E-types into three groups based on the inferred composition. The three groups are: "Nysa-like" E-types, whose spectra exhibit absorption bands at 0.9 and 1.8 μm, consistent with a mixture of aubrite plus olivine and orthopyroxene; "Angelina-like" asteroids, whose spectra show absorption bands at 0.5 and 0.9 μm, attributed to aubrite plus oldhamite; and "Hungaria-like" E-types, where only the absorption band at 0.9 μm is present, consistent with aubrite and possibly olivine.

In Figure 5 we plot the spectrum of 2015 TC25 along with the spectra of asteroids (44) Nysa, (64) Angelina, and (434) Hungaria, which represent each class under the classification derived by Clark et al. (2004). The NIR spectra of (44) Nysa and (64) Angelina were obtained as part of a dedicated program for the study of main belt asteroids. Asteroid (44) Nysa was observed on November 18, 2002, and (64) Angelina on August 14, 2003. The NIR spectra of both objects were obtained with the IRTF and SpeX, and the data reduction was carried out using Spextool. The spectrum of (434) Hungaria was obtained from Clark et al. (2004).

At first glance, the most obvious difference between the spectrum of 2015 TC25 and the other asteroids is the spectral slope. The spectrum of 2015 TC25 shows a negative (blue) slope, while the spectra of Nysa, Angelina and Hungaria all show positive (red) slopes. Variations in the spectral slope can be attributed to several factors, including composition, grain size, the phase angle, and space weathering. In the case of 2015 TC25 the grain size is likely playing the major role. 2015 TC25 is a small object with a very weak gravity field, which is also rotating very fast (once every ~ 133 s), making it very difficult for this object to retain a regolith layer on its surface. Hence, the surface of 2015 TC25 most likely resembles a bare rock. Figure 6 shows an example of the effect of the grain size on the spectra of aubrite ALH 78113. As can be seen in this figure increasing the grain size dramatically decreases the spectral slope. To better comparison between the spectra of 2015 TC25 and the other E-type asteroids we removed the continuum from all asteroid spectra, eliminating any effects caused by differences in grain size.

In Figure 7 we compare the continuum-removed spectra of 2015 TC25 with those of (44) Nysa, (64) Angelina, and (434) Hungaria. Of these three asteroids, the spectrum of 2015 TC25 seems to be most similar to those of (44) Nysa and (434) Hungaria, in particular the intensity and position of the 0.9 μm absorption band. We compared the Band I centers and depths of the four asteroids to verify any compositional affinity. Band I center of 2015 TC25 is 0.905±0.003 μm with a Band I depth of 5.4±0.2%. In the case of (44) Nysa the absorption band is centered at 0.903±0.002 μm and depth of 6.5±0.1%. For asteroid (64) Angelina we found that the Band I center is 0.936±0.002 μm, and the Band I depth is 1.2±0.1%. (434) Hungaria, on the other hand, has a Band I centered at 0.928±0.012 μm, and a Band I depth of 4.4±0.1%. These values indicate that the Band I center of 2015 TC25 is most similar to that of (44) Nysa and its Band depth is in between those measured for (434) Hungaria and (44) Nysa.

Based on these results one can rule out 2015 TC25 as an E(I) subtype in the Gaffey and Kelley (2004) classification, since the spectra of these objects are essentially featureless. Regarding the E(II) subtypes, their spectra exhibit absorption bands at ~ 0.5 and 0.96 µm, which have been associated to the calcium sulfide mineral, oldhamite. The absorption band of 2015 TC25 centered at 0.905 µm suggests that this mineral is absent, or if it is, the amount is too low to shift the Band I center to longer wavelengths. This makes it unlikely that 2015 TC25 belongs to the E(II) subtype. Interestingly, the visible spectra of (64) Angelina and (434) Hungaria exhibit the 0.5 µm feature (Fornasier et al. 2008), and their Band I centers is > 0.92 µm. This led to Fornasier et al. (2008) to classify these two objects as E(II).

Since both (64) Angelina and (434) Hungaria are classified as E(II) the possibility that 2015 TC25 belongs to either the 'Angelina-like" or "Hungaria-like" E-types (Clark et al. classification) seems to be remote. This elimination process leads to the conclusion that 2015 TC25 is a "Nysa-like" E-type, consistent with spectral match with (44) Nysa (Figure 7), and their similar spectral band parameters. (44) Nysa was classified as an E(III) subtype by Fornasier et al. (2008), which is the only remaining possibility under the Gaffey and Kelley (2004) classification. If 2015 TC25 is indeed an E(III) this would imply that this asteroid originated from a parent body that experienced extensive reduction during the igneous processing (Gaffey and Kelley 2004).

*3.2.2. Compositional analysis*

Currently there are no laboratory spectral calibrations for constraining the composition of 2015 TC25. We carried out a linear mixing model where we used a maximum of three end members following the approach by Nedelcu et al. (2007) and Fornasier et al. (2008) to model the spectra of E-type asteroids. The selection of the end members was based on the meteorites and minerals that are normally associated with E-type asteroids (e.g., Cloutis et al. 1990; Gaffey et al. 1992; Burbine et al. 2002; Clark et al. 2004; Fornasier et al. 2008). For the aubrite component of our model, we selected the spectrum of the Bishopville enstatite achondrite (HOSERLab sample ID MJG 225). We used this sample because is the only one we found whose spectrum exhibits an absorption feature near 0.9 µm and a negative slope (see Figure 4). The negative slope is likely caused by the fact that the sample is unsorted and it probably contains grains with large sizes. The other endmembers are: orthopyroxene $Wo_1En_{87}Fs_{12}$ (RELAB sample ID c1pe30), forsterite $Fo_{92}$ (RELAB sample ID c1po76), oldhamite CaS (RELAB sample ID c1tb38), and the standard Spectralon (RELAB sample ID c1hl04) used as neutral phase to reduce the spectral contrast (Nedelcu et al. 2007).

For the linear mixing model all the spectra were interpolated to a common wavelength sampling. We developed a Python routine that returns the optimal values for the end members so that the sum of the squares of the differences between the reflectance of the areal mixture and the asteroid's spectrum ($\chi 2$) is minimized. We found that the best fit is obtained with a combination of only two end members, Bishopville and orthopyroxene, in a proportion of 93 and 7%, respectively. The best fit obtained is shown in Figure 8. We noticed that including oldhamite as the third endmember will produce a mismatch in the position of the absorption band, which is shifted to longer wavelengths. This is consistent with 2015 TC25 belonging to the E (III) subtype rather than the E (II), discussed in the previous section. The addition of forsterite will have a similar effect, i.e., shifting the band center to longer wavelengths. The use of the neutral component did not improve the fit.

## 4. SOURCE REGION

In this section, we discuss the origin of 2015 TC25. It is a 2-meter diameter body with semimajor axis, eccentricity, and inclination values (a, e, i) of (1.030 AU, 0.118, 3.639°), respectively. This places into the Apollo-class of the NEA population. Using its Earth-like orbit, combined with insights provided by the NEO models of Bottke et al. (2002) and Granvik et al. (2015), we can predict where this body came from, at least in terms of the primary small body reservoirs of the inner solar system to delivered it to NEA orbits.

According to the Bottke et al. (2002) model, 2015 TC25 likely escaped the innermost region of the main belt through the ν6 resonance (65% probability) or the adjacent intermediate source Mars-crossing region (35%). If we instead use the Granvik et al. (2015) model, which is a newer, more advanced model that accounts for high inclination asteroid sources, 2015 TC25 has a 23% probability of coming from the Hungaria asteroid region, a 74% chance of coming from the ν6 resonance, and a 3% chance of coming from the 3:1 mean motion resonance with Jupiter. The two most likely candidate source regions would therefore be the Hungarias and the innermost region of the main belt.

In order to choose between these two options, we need to first provide some context for this body. 2015 TC25 is an E-type asteroid, which means it is similar to the aubrite meteorite class as demonstrated in the previous sections. The spectroscopic signatures of the E-types match the characteristics of enstatite, a high-albedo iron-poor mineral that comprise much of the aubrite meteorites (Gaffey and Kelley, 2004). The high albedos of the E-type asteroids (> 40%) also make candidate source asteroids highly distinctive in both the Hungaria and main belt populations when we consider data from infrared surveys like WISE (Masiero et al. 2011; 2013).

It is also useful to point out that when searching for meteorite sources, it is good to start by looking at asteroid families. Collision evolution models indicate families tend to produce numerous meteoroid-sized bodies over time via a collisional cascade, which can then reach main belt "escape hatches" via the Yarkovsky effect (Bottke et al. 2005). Accordingly, families tend to dominate non-family sources in terms of meteoroid production across the main belt (Bottke et al. 2005; 2015).

This information is useful when examining the Hungaria population. It is dominated by numerous E-type asteroids, with the prominent Hungaria family providing most of the known population (e.g., Warner et al. 2009b; Milani et al. 2009). The Hungaria family has also been linked as the source of most aubrite meteorites (Cuk et al. 2014). On average, aubrites have longer cosmic ray exposure ages than any other stony meteorite class, and this makes them a good match to the fairly long dynamical pathways needed for meteoroids to escape the Hungaria region and reach Earth. Given that 2015 TC25 is nearly meteoroid-sized itself, one must consider the Hungaria family as a likely candidate source family. As a fallback, one could also consider the non-family Hungaria region population as well, which also contains numerous E-type asteroids.

For inner main belt asteroids, however, there are fewer source options capable of creating 2015 TC25 and delivering it to its observed orbit. Consider that 2015 TC25's orbit is broadly similar to that of (101955) Bennu, the target of the OSIRIS-REx mission. Bottke et al. (2015) found that Bennu's orbit is difficult to reach from the inner main belt unless the body in question entered the nu6 resonance or IMC region from fairly low inclinations (i <8°; Bottke et al. 2015). This suggests the most plausible main belt parent bodies would also have low inclinations. Unfortunately, according to an analysis by Nesvorny et al. (2015), there are no known main belt asteroid families associated with E-type asteroids.

The best remaining candidate source body in the inner main belt would be the 70 km diameter E-type asteroid (44) Nysa. It resides in the inner main belt at (a, e, i) = (2.42 AU, 0.149, 3.7°). Based on the arguments provided by Bottke et al. (2015) (see also Campins et al. 2010), it seems probable that a large population of meteoroid-sized fragments ejected from (44) Nysa in a cratering event would produce at least a few bodies capable of reaching the observed orbit of 2015 TC25. The issue then is to quantify the relative probability that the Hungaria asteroids (or background population) are the source of 2015 TC25 vs. a cratering event on (44) Nysa that produces no observed family. While this problem is interesting, its solution would probably require an extensive modeling effort, so we leave it for future work.

## 5. SUMMARY

We observed near-Earth asteroid 2015 TC25 during a close flyby of the Earth at a distance of 69,000 miles on Oct. 12, 2015. Our comprehensive physical characterization effort involving optical, near-IR and radar data sets sheds light on the objects rotational, compositional properties, meteorite affinities and source regions from which it was derived from in the main asteroid below. Based on our study here is what we determined:

- Based on the composite light curve of 2015 TC25 we determined that it is a fast rotator with a rotational period of about 2.23 min (133.8 seconds). The light curve displays an amplitude of 0.40 mag with the peaks asymmetric by approximately 0.08 mag, suggesting that the object has an irregular shape.

- Assuming 2015 TC25 as triaxial ellipsoid with axes a>b>c and rotating along the c-axis, we estimated the object elongation (a/b) (Binzel et al. 1989) using Equation 4 of Kwiatkowski et al. (2010) to be ~ 1.22 suggesting a moderately deformed the object.

- Radar circular polarization ratio SC/OC for 2015 TC25 (0.62-1.33) is consistent with the range of polarization ratios measured for known E-type asteroids of 0.74 – 0.97.

- Near-IR spectrum of 2015 TC25 shows an absorption band at 0.9 μm and a negative spectral slope (decreasing reflectance with increasing wavelength) beyond 1.1 μm. The overall shape of the spectrum and band parameters of 2015 TC25 are consistent with E-type asteroids such as (44) Nysa, implying that this asteroid originated from a parent body that experienced extensive reduction during the igneous processing (Gaffey and Kelley 2004).

- Dynamical evidence suggests that the Hungaria family could be a source region for this asteroid, although it is also possible that a large population of meteoroid-sized fragments ejected from (44) Nysa in a cratering event would produce at least a few bodies capable of reaching the observed orbit of 2015 TC25.

This research work was supported by the NASA Near-Earth Object Observations Program grant NNX14AL06G (PI: Reddy). We thank the IRTF TAC for awarding time to this project, and we thanks the IRTF TOs and MKSS staff for their support. The IRTF is operated by the University of Hawaii under Cooperative Agreement no. NCC 5-538 with the National Aeronautics and Space Administration, Office of Space Science, Planetary Astronomy Program. Part of this work was done at the Arecibo Observatory, which is operated by SRI International under a cooperative agreement with the National Science Foundation (AST-1100968) and in alliance with Ana G. Mendez-Universidad Metropolitana and the Universities Space Research Association. The Arecibo Planetary Radar Program is supported by the National Aeronautics and Space Administration under Grant Nos. NNX12AF24G and NNX13AQ46G issued through the Near Earth Object Observations program.

**Tables**

**Table 1:** Observational circumstances for 2015 TC25.

| Date UTC | Telescope | Phase Angle (°) | V. Magnitude | Heliocentric Distance (AU) |
|---|---|---|---|---|
| Oct. 12, 2015 | DCT | 30 | 17.8 | 1.00 |
| Oct. 12, 2015 | MRO | 31 | 17.7 | 1.00 |
| Oct. 12, 2015 | IRTF | 32 | 17.5 | 1.00 |
| Oct. 17, 2015 | Arecibo | 150 | 28.0 | 0.99 |
| Oct. 18, 2015 | Arecibo | 149 | 28.0 | 0.99 |

**Figure Captions**

**Figure 1.** Light curve of 2015 TC25 obtained using the NASA IRTF (triangle); Magdalena Ridge Observatory (stars); and Discovery Channel Telescope (diamonds) showing a fast rotator with a rotational period of about 2.23 min (133.8 seconds). The composite lightcurve displays an amplitude of 0.40 mag with the peaks asymmetric by approximately 0.08 mag, suggesting that the object has an irregular shape.

**Figure 2.** Power spectrum of 2015 TC25 from Arecibo radar showing Doppler Frequency on the X-axis and echo power on the Y-axis. The polarization ratio SC/OC, determined as the quotient of the integrated SC (dotted line) and OC (bold line) echo powers in Figure 2, was 1.10 ± 0.23 and 0.77 ± 0.15 (one-sigma) on October 17 (2A) and 18 (2B), respectively. Combining the observations suggests a polarization ratio of ~ 0.9 with a possible range of 0.62 – 1.33.

**Figure 3.** Near-IR spectrum of 2015 TC25 obtained using the NASA Infrared Telescope Facility on Mauna Kea, Hawai'i. The spectrum has been normalized to unity at 0.8 µm. This spectrum shows a weak absorption band at 0.9 µm and a negative spectral slope (decreasing reflectance with increasing wavelength) beyond 1.1 µm.

**Figure 4.** Reflectance spectra of aubrites Peña Blanca (RELAB sample ID c1tb45), Happy Canyon (RELAB sample ID cbea03), Mayo Belwa (RELAB sample ID s1tb46), ALH 78113 (RELAB sample ID bkr1ar001B), and Bishopville (HOSERLab sample ID MJG 225).

**Figure 5.** The spectrum of 2015 TC25 along with the spectra of main belt E-type asteroids (44) Nysa, (64) Angelina, and (434) Hungaria. Spectra of (44) Nysa and (64) Angelina were obtained by us using the NASA IRTF. The spectrum of (434) Hungaria is from Clark et al. (2004). Spectra are normalized to unity at 0.8 µm and are offset for clarity.

**Figure 6.** A plot showing the effect of the grain size on the spectra of aubrite ALH 78113. Increasing the grain size dramatically decreases the spectral slope.

**Figure 7.** A plot comparing the continuum-removed spectra of 2015 TC25 with those of (44) Nysa, (64) Angelina, and (434) Hungaria. Of these three asteroids, the spectrum of 2015 TC25 is similar to those of (44) Nysa and (434) Hungaria, in particular the intensity and position of the 0.9 µm absorption band.

**Figure 8.** A plot comparing the spectrum of 2015 TC25 with the results of our linear mixing model following the approach by Nedelcu et al. (2007) and Fornasier et al. (2008) to model the spectra of E-type asteroids. We found that the best fit is obtained with a combination of only two end members, Bishopville aubrite and orthopyroxene, in a proportion of 93 and 7%, respectively.

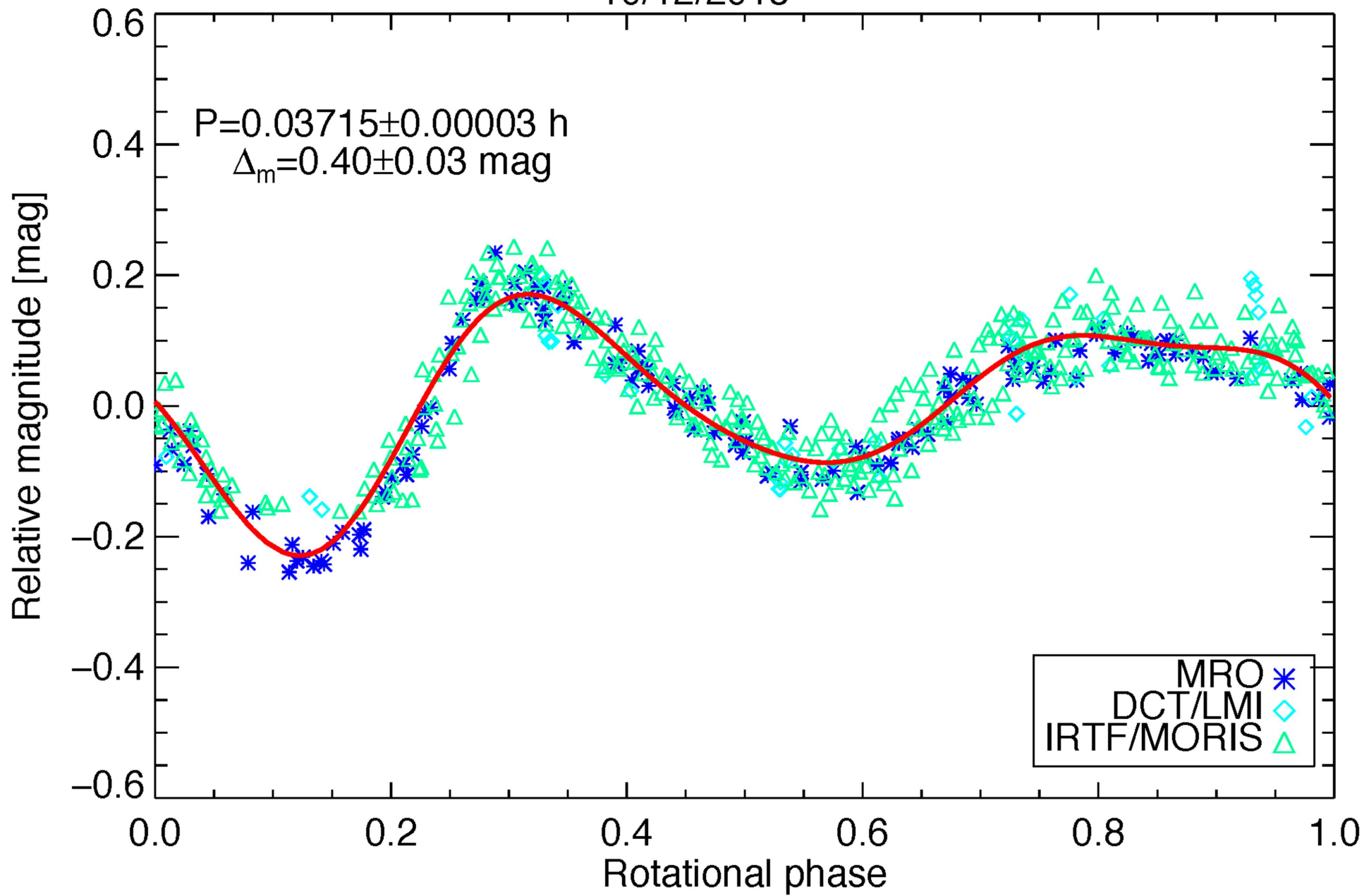

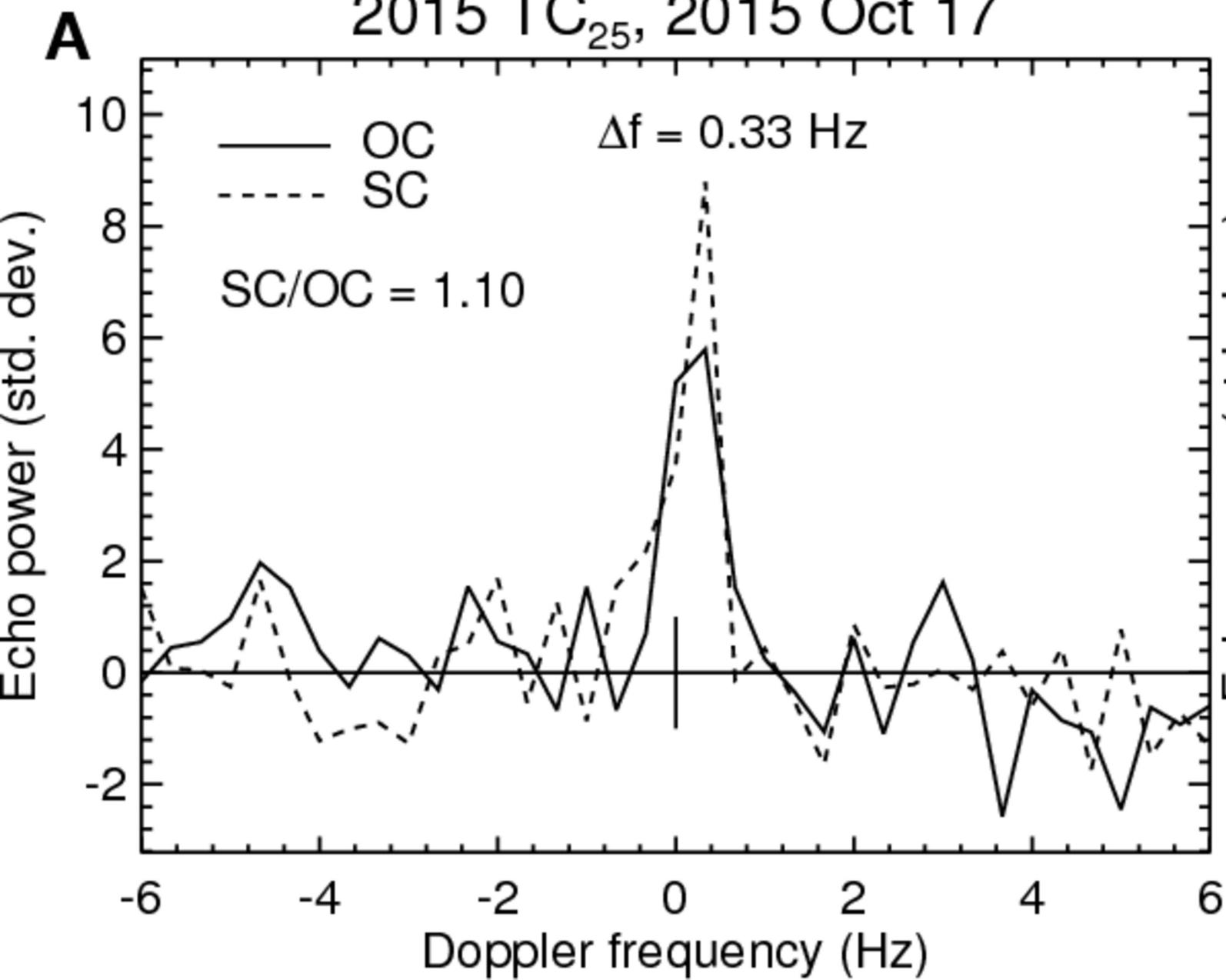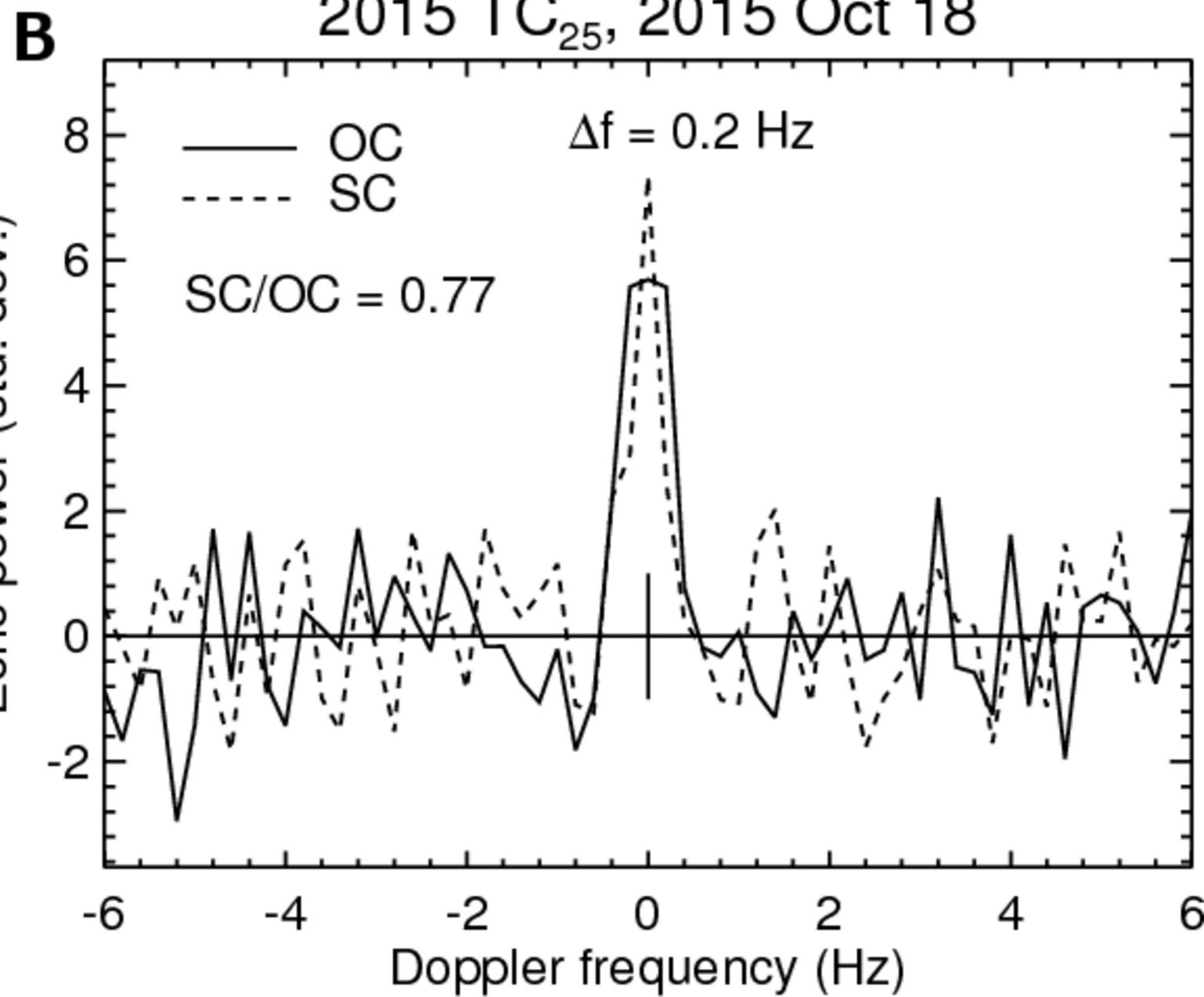

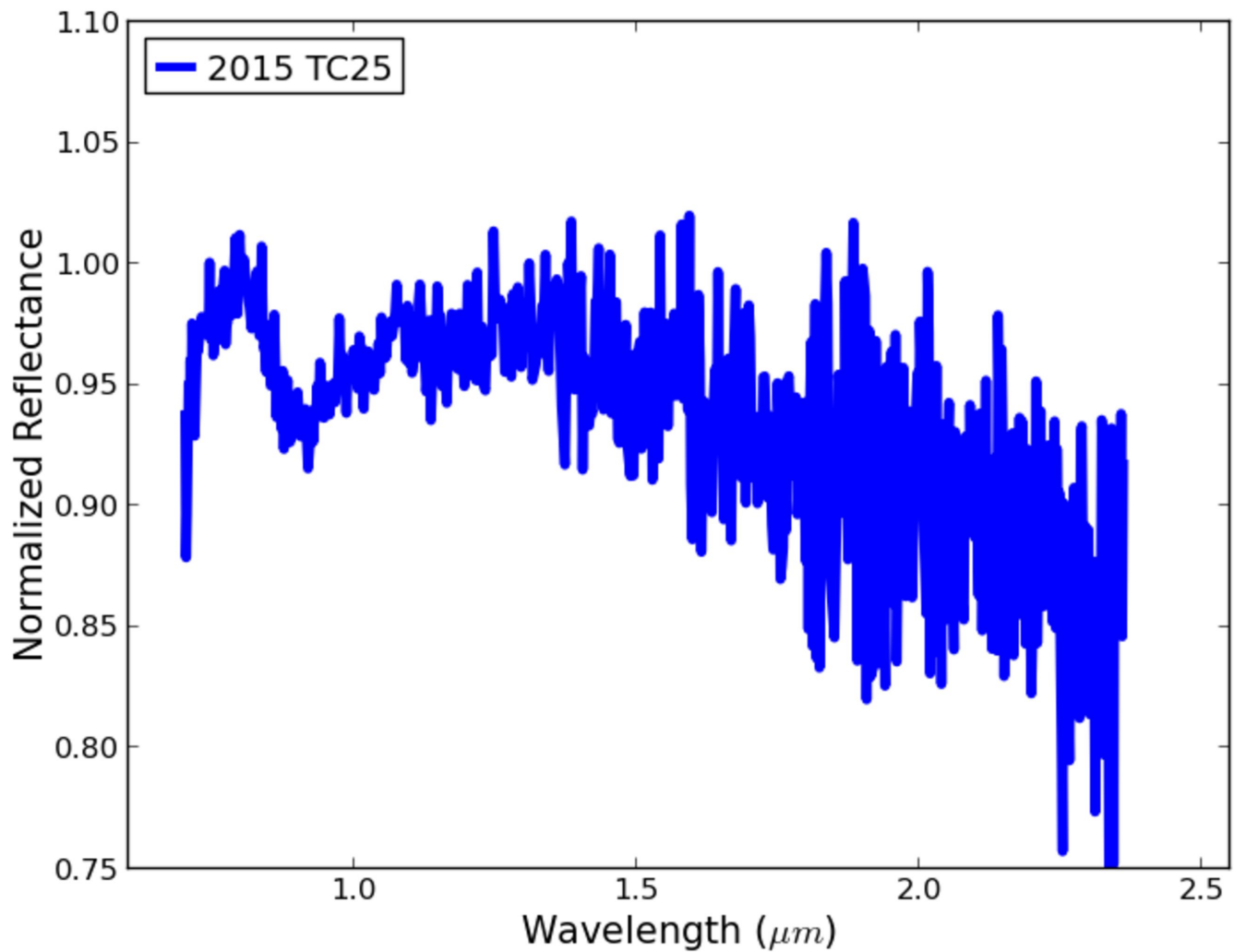

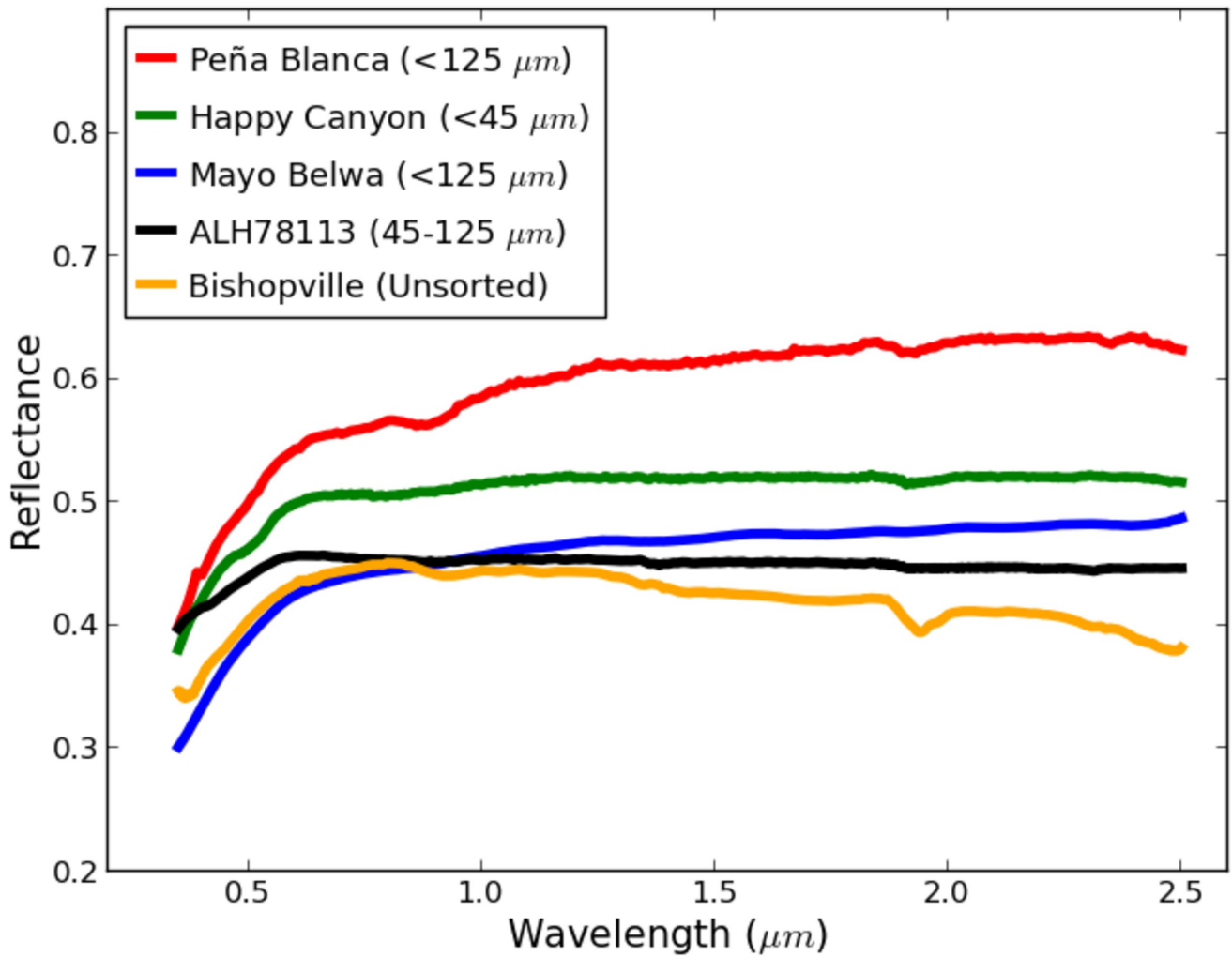

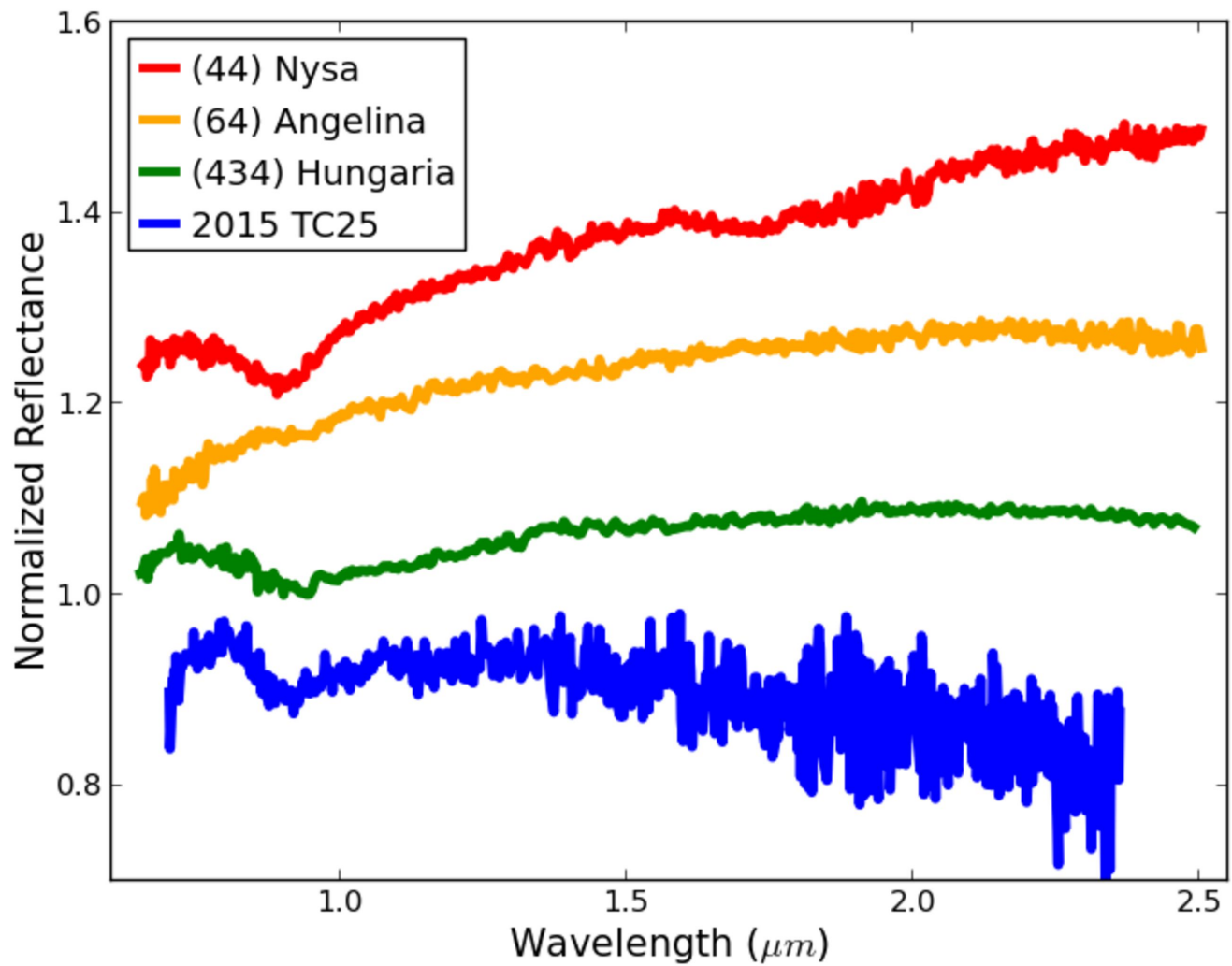

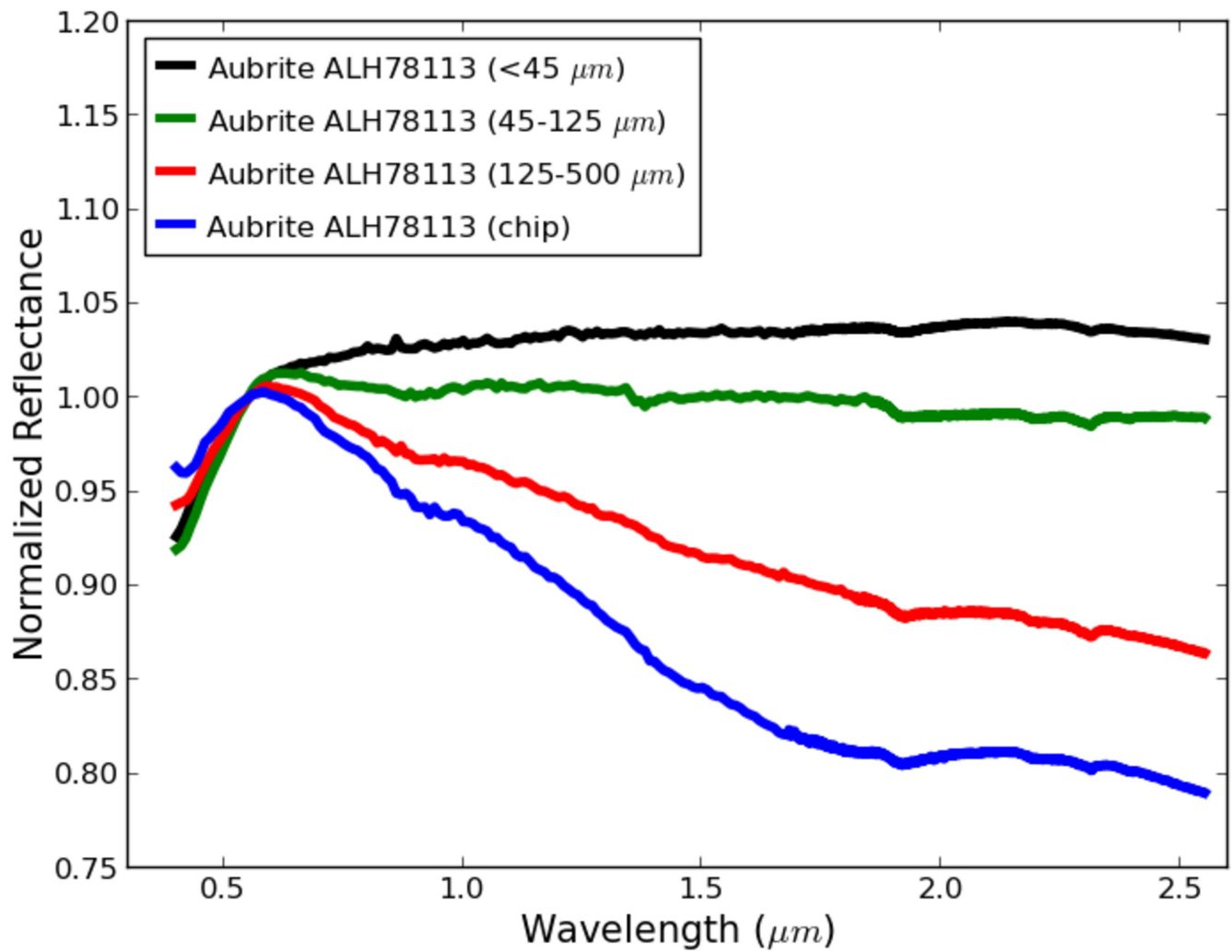

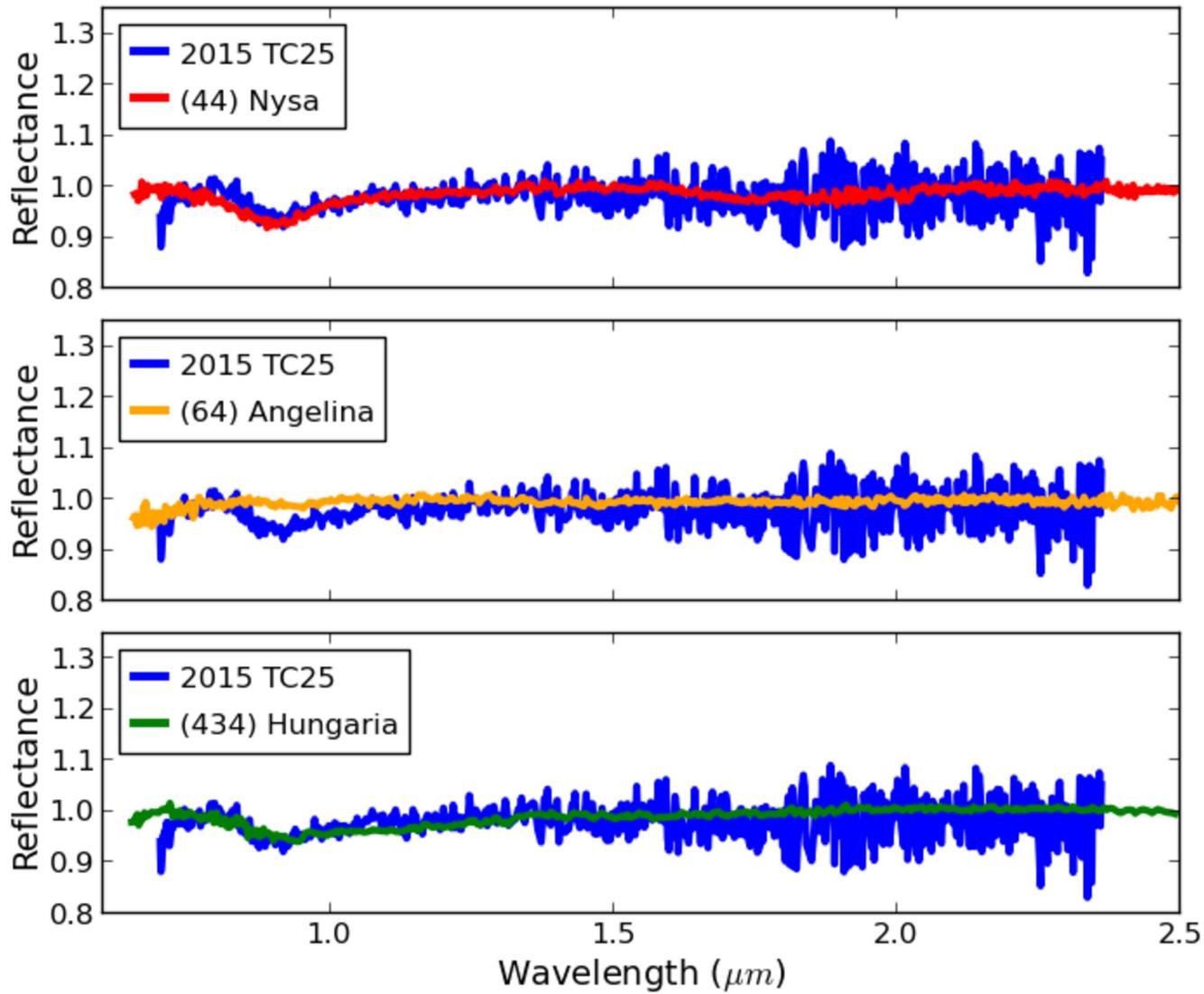

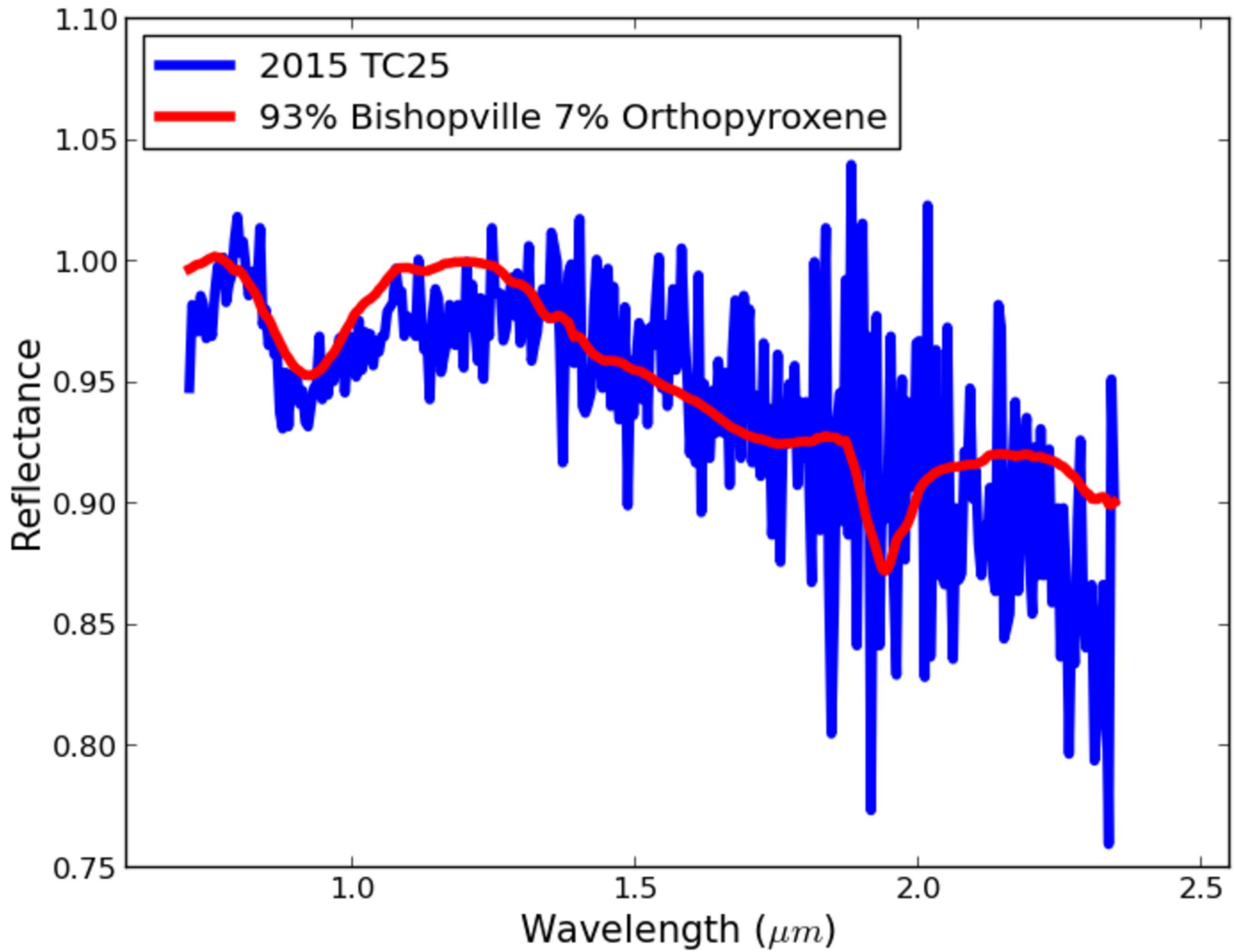